\documentclass[%
 reprint,
 superscriptaddress,
 nofootinbib,
 amsmath,amssymb,
 aps,
floatfix
]{revtex4-1}
\makeatletter
\newcommand*{\rom}[1]{\expandafter\@slowromancap\romannumeral #1@}
\makeatother

\usepackage[english]{babel}
\usepackage[dvipsnames]{xcolor}
\usepackage{titlesec}
\usepackage{graphicx}
\usepackage{epsfig}
\usepackage{amsmath,epsfig}
\usepackage{bm}
\usepackage{amssymb}
\usepackage{subfig}
\usepackage{appendix}
\usepackage{graphicx}
\usepackage{dcolumn}
\usepackage{bm}
\usepackage{txfonts}
\usepackage[utf8]{inputenc}
\newcommand{\be}{\begin{equation}}
\newcommand{\ee}{\end{equation}}
\newcommand{\ba}{\begin{eqnarray}} 
\newcommand{\ea}{\end{eqnarray}}

\newcommand{\beq}{\begin{equation}}
\newcommand{\eeq}{\end{equation}}
\newcommand{\bea}{\begin{eqnarray}}
\newcommand{\eea}{\end{eqnarray}}
\newcommand{\beqa}{\begin{eqnarray}}
\newcommand{\eeqa}{\end{eqnarray}}
\newcommand{\bseq}{\begin{subequations}}
\newcommand{\eseq}{\end{subequations}}

\definecolor{M_Green}        {rgb}{0.03 , 0.6 , 0.35}

\def\0{{\boldsymbol 0}}

\begin{document}

\title{Accelerating longitudinal expansion of resistive relativistic-magneto-hydrodynamics  in heavy ion collisions}

\author{M.~Haddadi Moghaddam}
\email{haddadi@to.infn.it \& hadadi$_$65@yahoo.com}
\affiliation{Department of Physics, University of Turin and INFN,
\small Turin
Via Pietro Giuria 1, I-10125 Turin, Italy}

\author{W.~M.~Alberico}
\affiliation{Department of Physics, University of Turin and INFN,
\small Turin
Via Pietro Giuria 1, I-10125 Turin, Italy}

\author{Duan She}
\affiliation{Key Laboratory of Quark and Lepton Physics, Ministry of Education, Wuhan, 430079, China}
\affiliation{Institute of Particle Physics, central China Normal University, Wuhan 430079, China}
\affiliation{Physics Department and Center for Exploration of Energy and Matter,
Indiana University, 2401 N Milo B. Sampson Lane, Bloomington, Indiana 47408, USA}

\author{A.~F.~Kord}
\affiliation{Department of physics, Hakim Sabzevari University (HSU),
\small P.O.Box 397, Sabzevar, Iran}

\author{B.~Azadegan}
\affiliation{Department of physics, Hakim Sabzevari University (HSU),
\small P.O.Box 397, Sabzevar, Iran}


\begin{abstract}
We study the evolution of the longitudinal expansion of an ideal fluid with finite electrical conductivity, which is subject to the EM fields. In the framework of resistive relativistic-magneto-hydrodynamic, we find an
 exact analytical solution for the EM fields and for the acceleration of the fluid.

\end{abstract}

\maketitle

\section{Introduction}
Relativistic heavy ion collisions, provide an opportunity to study the matter produced in a de-confined partons state during the collisions. This matter is called quark gluon plasma (QGP), and lives for life times of the order of some $fm$s. The aforementioned matter has been successfully described within the relativistic hydrodynamic frame work (\cite{Romatschke}-\cite{delzanna2013}). In a simple scenario, the transverse expansion of the fluid is neglected and the longitudinal expansion is considered within the well-known Bjorken model\cite{bj83}. However, in a more realistic world, the longitudinal expansion could be affected by acceleration: as a consequence there are boost non-invariant initial conditions and there need not to be a rapidity plateau (\cite{eskola1998}-\cite{bozek2008}). A recent description based on accelerating hydrodynamic description can be found in Ref.\cite{Csanad2018}.

Recently, it was argued that due to the charged particle motion in Relativistic Heavy Ion Collisions (RHIC), huge electromagnetic (EM) fields are produced: in the energy range of interest in RHIC, the strength of this EM fields are $e\mid B\mid/m_\pi^2\approx 1-3$ up to $e\mid B\mid/m_\pi^2\approx 10-15$ at the LHC (Large Hadron Collider) energies \cite{Tuchin:2013apa}-\cite{voronyuk}. In general, these fields decay very fast, but the existence of a QGP fluid with electrical conductivity leads them to be more steady in time. Hence, it might be possible to detect the effects of EM fields on the observables as well as a variety of phenomena like chiral magnetic effect and chiral magnetic wave \cite{kharzeev2008}-\cite{kharzeevprl}.

Now, the presence of EM fields in relativistic heavy ion collisions requires to solve the problem, e.g., in the context of relativistic magneto-hydrodynamics (RMHD). Several papers are present in the literature, which perform analytical and numerical calculations within RMHD, by assuming infinite electrical conductivity  (Ideal RMHD) \cite{roy15}-\cite{duanshe2019}. However, it has been deduced from lattice QCD calculations \cite{ding2011}-\cite{amato2013} that the electrical conductivity of the plasma under investigation corresponds to the ambient temperature and that a finite value could be more appropriate. Hence, we will consider here the Resistive Relativistic Magneto-Hydrodynamic (RRMHD) framework.

In this paper, assuming a finite electric conductivity for the quark gluon plasma, the longitudinal motion of the plasma embedded into electromagnetic fields is studied. In this way, the Bjorken flow is generalized and the solutions found through RRMHD are not necessarily  Lorentz invariant.
Moreover it is assumed that the electromagnetic fields are oriented in the transverse plane and are perpendicular to the plasma velocity: this is called {\it transverse MHD}; also, we do not take into account any vorticity. Finally,  analytical solutions for the longitudinal fluid evolution as well as for the electromagnetic fields are presented.
In this study, it is shown that due to the presence of electromagnetic fields, the energy density of the fluid decreases faster with time than in the Bjorken model and therefore it appears that the energy flows towards high rapidity.

The paper is organized as follows: in Sec.  \text{II} we present the RRMHD framework. Sec. \text{III}  is dedicated to the (1+1) longitudinal expansion of RRMHD. Finally in Sec. \text{IV} we discuss the results and draw our conclusions.
\section{Resistive relativistic magneto-hydrodynamic}
In order to describe the interaction of matter and electromagnetic fields in the quark-gluon  plasma we consider the relativistic magneto-hydrodynamics (RMHD)  framework \cite{geod}-\cite{an89}. For the sake of simplicity, we assume an ideal
relativistic plasma  with  massless particles and  finite electrical conductivity ($\sigma$).
In addition, the fluid is considered  to be ultra relativistic, thus implying that the rest-mass
contributions to the equation of state (EOS) have been neglected, and the pressure is simply
proportional to the energy density: $P = \kappa\epsilon$ where $\kappa$ is constant.
 For an ideal fluid with finite electrical conductivity, which is called resistive fluid, the equations of RMHD can
be written in the form of the covariant conservation laws
\begin{eqnarray}
d_\mu T_{matter}^{\mu\nu}&=&-J_\lambda F^{\lambda\nu} ,\label{1}\\
d_\mu F^{\star\mu\nu}&=&0,\label{2}\\
d_\mu F^{\mu\nu}&=&-J^\nu,\ d_\mu J^\mu=0 \label{3}
\end{eqnarray}
Where $d_\mu$ is the covariant derivative and the energy momentum tensor for the fluid is
\begin{eqnarray}
T_{matter}^{\mu\nu}&=&(\epsilon+P)u^\mu u^\nu+Pg^{\mu\nu},\  P=\kappa\epsilon,
\end{eqnarray}
 $u^\mu$,$ \epsilon$ and $P$ are the fluid four velocity, energy density and pressure, respectively.
The electromagnetic field tensors and the current density are given by:
\begin{eqnarray}
F^{\mu\nu}&=&u^\mu e^\nu-u^\nu e^\mu+\epsilon^{\mu\nu\lambda\kappa}b_\lambda u_\kappa,\\
F^{\star\mu\nu}&=&u^\mu b^\nu-u^\nu b^\mu-\epsilon^{\mu\nu\lambda\kappa}e_\lambda u_\kappa,\\
J^\mu&=&\rho u^\mu+\sigma e^\mu
\end{eqnarray}
 Where $\rho$ is the proper charge density and $\epsilon^{\mu\nu\lambda\kappa}=(-g)^{-1/2}[\mu\nu\lambda\kappa]$ is the space time Levi-Civita tensor density ($\epsilon_{\mu\nu\lambda\kappa}=-(-g)^{1/2}[\mu\nu\lambda\kappa]$) with $g=\det\{g_{\mu\nu}\}$ and $[\mu\nu\lambda\kappa]$ is the alternating Levi-Civita symbol\footnote{$[\mu\nu\lambda\kappa]$ is the totally anti-symmetric symbol defined as:
 
 $[\mu\nu\lambda\kappa]:= \begin{cases}
    1 & \text{for any even permutation of 0, 1, 2, 3,} \\
    -1 & \text{for odd permutations of 0, 1, 2, 3,} \\
    0 & \text{for any case with repeated indices.}
  \end{cases}$}.
Besides:
\begin{eqnarray}
e^\mu=F^{\mu\nu}u_\nu,\ b^\mu=F^{\star\mu\nu}u_\nu, \  (e^\mu u_\mu=b^\mu u_\mu=0)
\end{eqnarray}
$e^\mu$ and  $b^{\mu}$ being the electric and  magnetic field four
vectors in the co-moving frame of the  fluid, which is related to the one measured in the lab-frame.
 Moreover,  the fluid  four velocity $u_{\mu}$ ($u_{\mu}u^{\mu}=-1$) is given by:
$$u^\mu=\gamma(1, \vec v),\ \quad \gamma=\frac{1}{\sqrt{1-v^2}}$$
In eqs.(\ref{1}) to (\ref{3}) the covariant derivatives are given by:
\begin{eqnarray}
d_\mu A^\nu&=&\partial_\mu A^\nu+\Gamma^\nu_{\mu m} A^m\\
d_p A^{\mu\nu}&=&\partial_p A^{\mu\nu}+\Gamma^\mu_{p m} A^{m \nu}+\Gamma^\nu_{p m} A^{m \mu},
\end{eqnarray}
where $\Gamma^i_{j k}$ are the Christoffel symbols
\begin{eqnarray}
\Gamma^i_{jk}=\frac{1}{2}g^{im}\left(\frac{\partial g_{mj}}{\partial x^k}+\frac{\partial g_{mk}}
{\partial x^j}-\frac{\partial g_{jk}}{\partial x^m}\right)
\end{eqnarray}

Hereafter, instead of  the standard Cartesian coordinates it is preferable to use Milne coordinates for a longitudinal flow:
\begin{eqnarray}
(\tau, x, y, \eta)&=&\left(\sqrt{t^2-z^2},x,y,\frac{1}{2}ln\frac{t+z}{t-z}\right).
\end{eqnarray}
Here, the metric\footnote{In Cartesian coordinates, the metric tensor $g_{\mu\nu}$ of the special-relativistic space time, known as the Minkowski space time, is simply given by $g_{\mu\nu}=g^{\mu\nu}=diag(-1, 1, 1, 1)$.} is given by:
\begin{eqnarray}
g^{\mu\nu}=diag(-1, 1, 1, 1/\tau^2),  \  \  \  \ g_{\mu\nu}=diag(-1, 1, 1, \tau^2).
\end{eqnarray}

Working in  Milne coordinates, one can easily obtain the Christoffel symbols: the only non-vanishing ones
being: $\Gamma^\tau_{\eta\eta}=\tau$ and $\Gamma^\eta_{\eta\tau}=\Gamma^\eta_{\tau\eta}=1/\tau$.

By implementing the projection of $d_\mu T_{matter}^{\mu\nu}=-J_\lambda F^{\lambda\nu}$ along the longitudinal and transverse directions with respect to $u^\mu$, one can rewrite the conservation equations as
\begin{eqnarray}
&&u_\nu( d_\mu T_{matter}^{\mu\nu}=-J_\lambda F^{\lambda\nu})
\rightarrow D\epsilon+(\epsilon+P)\Theta=e^\lambda J_\lambda,\label{e}\\
&&\Delta^\alpha_\nu ( d_\mu T_{matter}^{\mu\nu}=-J_\lambda F^{\lambda\nu})\nonumber\\
&&\quad\rightarrow(\epsilon+P)Du^\alpha+\nabla^\alpha P=
g^\alpha_\nu F^{\nu\lambda}J_\lambda-u^\alpha e^\lambda J_\lambda. \label{m}
\end{eqnarray}
Where
\begin{eqnarray}
&D=u^\mu d_\mu,\ \Theta=d_\mu u^\mu,\ \nabla^\mu=d^\mu+u^\mu D,
\nonumber\\
&\qquad\Delta^\alpha_\nu=g^\alpha_\nu+u^\alpha u_\nu
\end{eqnarray}

\section{(1+1) Longitudinal expansion with acceleration}
Here we assume, during the whole evolution, that the velocity of the fluid is directed in the longitudinal direction, while the transverse flow is neglected.
Hence we can parameterize the fluid four-velocity in (1+1D) as follows
\begin{eqnarray}
  u^\mu=\gamma(1,0, 0, v_z)=(\cosh Y, 0, 0, \sinh Y),
\end{eqnarray}
where $Y$ is the fluid rapidity and $v_z=\tanh Y$. Besides, in Milne coordinates, one can write
\begin{eqnarray}
  u^\mu&&= \left(\cosh (Y-\eta), 0, 0, \frac{1}{\tau}\sinh (Y-\eta)\right)
  \nonumber\\
  &&=\bar{\gamma}(1, 0, 0, \frac{1}{\tau}\bar{v}),
\end{eqnarray}
where
\begin{eqnarray}
  \bar{\gamma}=\cosh (Y-\eta),\ \bar{v}=\tanh(Y-\eta).
\end{eqnarray}
By using this parameterization  one obtains
\begin{eqnarray}
  D&=&\bar{\gamma}(\partial_\tau+\frac{1}{\tau}\bar{v}\partial_\eta)  \\
  \Theta &=& \bar{\gamma}(\bar{v}\partial_\tau Y+\frac{1}{\tau}\partial_\eta Y)
\end{eqnarray}
Now, Eq. (\ref{e}) leads to
\begin{eqnarray}
 (\tau\partial_{{\tau}}+\bar{v}\partial_\eta)\epsilon+(\epsilon+P)(\tau\bar{v}\partial_{{\tau}} Y+\partial_\eta Y) &=& \bar{\gamma}^{(-1)}\tau \ e^\lambda J_\lambda,\label{energy}
\end{eqnarray}
and Eq. (\ref{m}), for $\alpha=\eta$, gives
\begin{eqnarray}
  (\epsilon+P)D u^\eta+\nabla^\eta P &=& F^{\eta \lambda}J_\lambda -u^\eta (e^\lambda J_\lambda)
\end{eqnarray}
with
\begin{eqnarray}
  D u^\eta &=& \frac{1}{\tau^2}\bar{\gamma}^2(\tau\partial_{{\tau}}+\bar{v}\partial_\eta)Y  \\
  \nabla^\eta P &=& \frac{1}{\tau^2}\bar{\gamma}^2 (\tau\bar{v}\partial_{{\tau}} +\partial_\eta )P.
  \end{eqnarray}
  Finally, for the Euler equation (\ref{m}) we get the expression:
\begin{eqnarray}
  &&(\epsilon+P)(\tau\partial_{{\tau}}+\bar{v}\partial_\eta)Y+(\tau\bar{v}\partial_{{\tau}} +\partial_\eta )P =
  \nonumber\\
  &&\qquad\quad =
  \bar{\gamma}^{(-2)}\tau^2 \Big[ F^{\eta \lambda}J_\lambda -u^\eta e^\lambda J_\lambda \Big],\label{euler}
\end{eqnarray}

In general,  the electric and magnetic four vectors are considered in the transverse plane  as follows
\begin{eqnarray}
  e^\mu = (0, e^x, e^y, 0),\\ b^\mu=(0, b^x, b^y, 0)
\end{eqnarray}
Where $e, \ b$ are the magnitudes of the EM fields.
Then the relevant components of the electromagnetic tensor and induced current are
$$F^{\eta x}=\frac{\bar{\gamma}}{\tau}(b^y+\bar{v} e^x),\ J^x=\sigma e^x $$
$$F^{\eta y}=\frac{\bar{\gamma}}{\tau}(-b^x+\bar{v} e^y),\ J^y=\sigma e^y $$
Where $u^\mu e_\mu= u^\mu b_\mu=0$ is satisfied. So, Eqs. (\ref{energy}) and (\ref{euler}) become
\begin{eqnarray}
  (\tau\partial_{{\tau}}+\bar{v}\partial_\eta)\epsilon+(\epsilon+P)(\tau\bar{v}\partial_{{\tau}} Y+\partial_\eta Y) &=& \bar{\gamma}^{(-1)}\tau \sigma\ (e_x^2+e_y^2),
  \nonumber\\ \label{eq1} \\
  (\epsilon+P)(\tau\partial_{{\tau}}+\bar{v}\partial_\eta)Y+(\tau\bar{v}\partial_{{\tau}} +\partial_\eta )P &=&\bar{\gamma}^{(-1)}\tau \sigma\ (e^x b^y-e^y b^x)\nonumber\\ \label{eq2}
\end{eqnarray}

Clearly, due to the EM fields, the boost invariance of the solutions is broken.
However, if electric and magnetic field would be in parallel or anti-parallel directions (\cite{shokri2017}-\cite{irfan2019}) (it is not the case in Heavy Ion Collisions setup), then the r.h.s of the Euler equation (\ref{eq2}) will disappear. By considering a  time dependent evolution for the medium, the fluid under study will not accelerate and Bjorken flow is preserved.

 However, non central  collisions can create an out-of-plane magnetic field and in-plane electric field. The magnetic field in non central
 collisions is dominated by the $y$ component which induces a Faraday
 current in $xz$ plane (\cite{catania2017}- \cite{gursoy2018}). 
 In particular, we are here interested in obtaining   solutions
 representing the RMHD extension of one-dimensional generalized
 Bjorken flow ($v_z\neq \frac{z}{t}$) along the z-direction with
 velocity $u^\mu=\gamma(1, 0, 0, v_z)$, the Lorentz force being
 directed along the $x$ direction.  We assume that the electric field
 is oriented in $x$ direction and   the magnetic field is
 perpendicular to the reaction plane, pointing along the y direction
 in an in-viscid fluid with finite electrical conductivity, flow
 expansion being along the $z$ direction.

The homogeneous Maxwell equation, $d_\mu F^{*\mu\nu}=0$,
 lead to the following equations:
\begin{eqnarray}
\partial_x F^{*x\tau}+\partial_y F^{*y\tau}+\partial_\eta F^{*\eta\tau}&=&0,\label{M1}\\
\partial_\tau F^{*\tau x}+\partial_y F^{*y x}+\partial_\eta F^{*\eta x}+\frac{1}{\tau}F^{*\tau x}&=&0,\label{M2}\\
\partial_\tau F^{*\tau y}+\partial_x F^{*x y}+\partial_\eta F^{*\eta y}+\frac{1}{\tau}F^{*\tau y}&=&0,\label{M3}\\
\partial_\tau F^{*\tau \eta}+\partial_x F^{*x\eta}+\partial_y F^{*y \eta}+\frac{1}{\tau}F^{*\tau\eta}&=&0.\label{M4}
\end{eqnarray}
In the same way, the in-homogeneous Maxwell equations $d_\mu F^{\mu\nu}=-J^\nu$ are given by:
\begin{eqnarray}
\partial_x F^{x\tau}+\partial_y F^{y\tau}+\partial_\eta F^{\eta\tau}&=&-J^\tau,\label{IM1}\\
\partial_\tau F^{\tau x}+\partial_y F^{y x}+\partial_\eta F^{\eta x}+\frac{1}{\tau}F^{\tau x}&=&-J^x,\label{IM2}\\
\partial_\tau F^{\tau y}+\partial_x F^{x y}+\partial_\eta F^{\eta y}+\frac{1}{\tau}F^{\tau y}&=&-J^y,\label{IM3}\\
\partial_\tau F^{\tau \eta}+\partial_x F^{x\eta}+\partial_y F^{y \eta}+\frac{1}{\tau}F^{\tau\eta}&=&-J^\eta,\label{IM4}
\end{eqnarray}
Let's now consider the following setup:
\begin{eqnarray}\label{setup}
  u^\mu= (\cosh (Y-\eta), 0, 0, \frac{1}{\tau}\sinh (Y-\eta)),\nonumber\\
   e^\mu = (0, e^x, 0, 0),\ b^\mu=(0, 0, b^y, 0).
\end{eqnarray}
After substituting the above setup (\ref{setup})  in Maxwell equations we obtain:
\begin{eqnarray}
 && \partial_\tau\left[(u^\tau b^y+\frac{1}{\tau}e_x u_\eta)\right]+\partial_\eta\left[(u^\eta b^y-\frac{1}{\tau}e_x u_\tau)\right]\nonumber\\
 &&\qquad +\frac{1}{\tau}\left[(u^\tau b^y+\frac{1}{\tau}e_x u_\eta)\right]=0,\label{max1}\\
 &&  \partial_\tau\left[(u^\tau e^x+\frac{1}{\tau} b_y u_\eta)\right]+\partial_\eta\left[(u^\eta e^x-\frac{1}{\tau}b_y u_\tau)\right]\nonumber\\
 &&\qquad +\frac{1}{\tau}\left[(u^\tau e^x+\frac{1}{\tau} b_y u_\eta)\right]=-\sigma e_x. \label{max2}
\end{eqnarray}
We suppose that all quantities are constant in the transverse plane.
Hence in order to solve the last two equations, we can write the following Ansatz\footnote{By considering this kind of Ansatz, we find that the EM fields in the lab frame would be as follow: $\bf{E}^{i}_{L}=F^{0 i}=0$ and ${\bf{B}^{i}_{L}=F^{\star 0 i}}\rightarrow B^y_L=h(\tau,\eta)$.}:
\begin{eqnarray}
  e_x(\tau,\eta) &=& -h(\tau, \eta) \sinh(Y-\eta)\label{ansatz1} \\
 b_y(\tau,\eta) &=& h(\tau, \eta) \cosh(Y-\eta)\label{ansatz2}
\end{eqnarray}
then Eqs. (\ref{max1}-\ref{max2}) give:
\begin{eqnarray}
  \partial_\tau h(\tau, \eta)+\frac{h(\tau, \eta)}{\tau} &=& 0, \label{h1} \\
  \partial_\eta h(\tau, \eta)+\sigma\tau h(\tau, \eta) \sinh(\eta-Y) &=& 0,\label{h2}
\end{eqnarray}
and the solution of Eq. (\ref{h1}) can be written as:
\begin{eqnarray}
  h(\tau, \eta) &=& \frac{c(\eta)}{\tau},
\end{eqnarray}
where,  $c(\eta)$ is an arbitrary function.

Moreover from eq. (\ref{h2}) we can find
\begin{eqnarray}
  \sinh(Y-\eta) &=& \frac{1}{\sigma\tau}\frac{\partial_\eta c(\eta)}{c(\eta)}
\end{eqnarray}
and
\begin{eqnarray}
  \cosh(Y-\eta) &=& \sqrt{1+\frac{1}{\sigma^2\tau^2}\left(\frac{\partial_\eta c(\eta)}{c(\eta)}\right)^2}
\end{eqnarray}

We summarize the solutions for fluid rapidity, four velocity profile and EM fields as follows:
\begin{eqnarray}
        Y&=& \eta+\sinh^{-1}\left(\frac{1}{\sigma\tau}\frac{\partial_\eta c(\eta)}{c(\eta)}\right),\\
        u^\tau &=& \sqrt{1+\frac{1}{\sigma^2\tau^2}\left(\frac{\partial_\eta c(\eta)}{c(\eta)}\right)^2},  \\
  u^\eta &=& \frac{1}{\sigma\tau^2}\frac{\partial_\eta c(\eta)}{c(\eta)},\\
  e_x(\tau, \eta) &=& - \frac{1}{\sigma\tau^2}\frac{\partial c(\eta)}{\partial\eta},  \\
  b_y(\tau, \eta) &=& \frac{c(\eta)}{\tau}\times \sqrt{1+\frac{1}{\sigma^2\tau^2}\left(\frac{\partial_\eta c(\eta)}{c(\eta)}\right)^2}.
\end{eqnarray}
The time dependence of the above solutions is clear but the $\eta$ dependence profile is not yet known. Indeed we do not get a unique solution for the system under investigation, unless we impose other restrictions on the solutions.


\begin{figure}
\begin{center}
\includegraphics[width=3in]{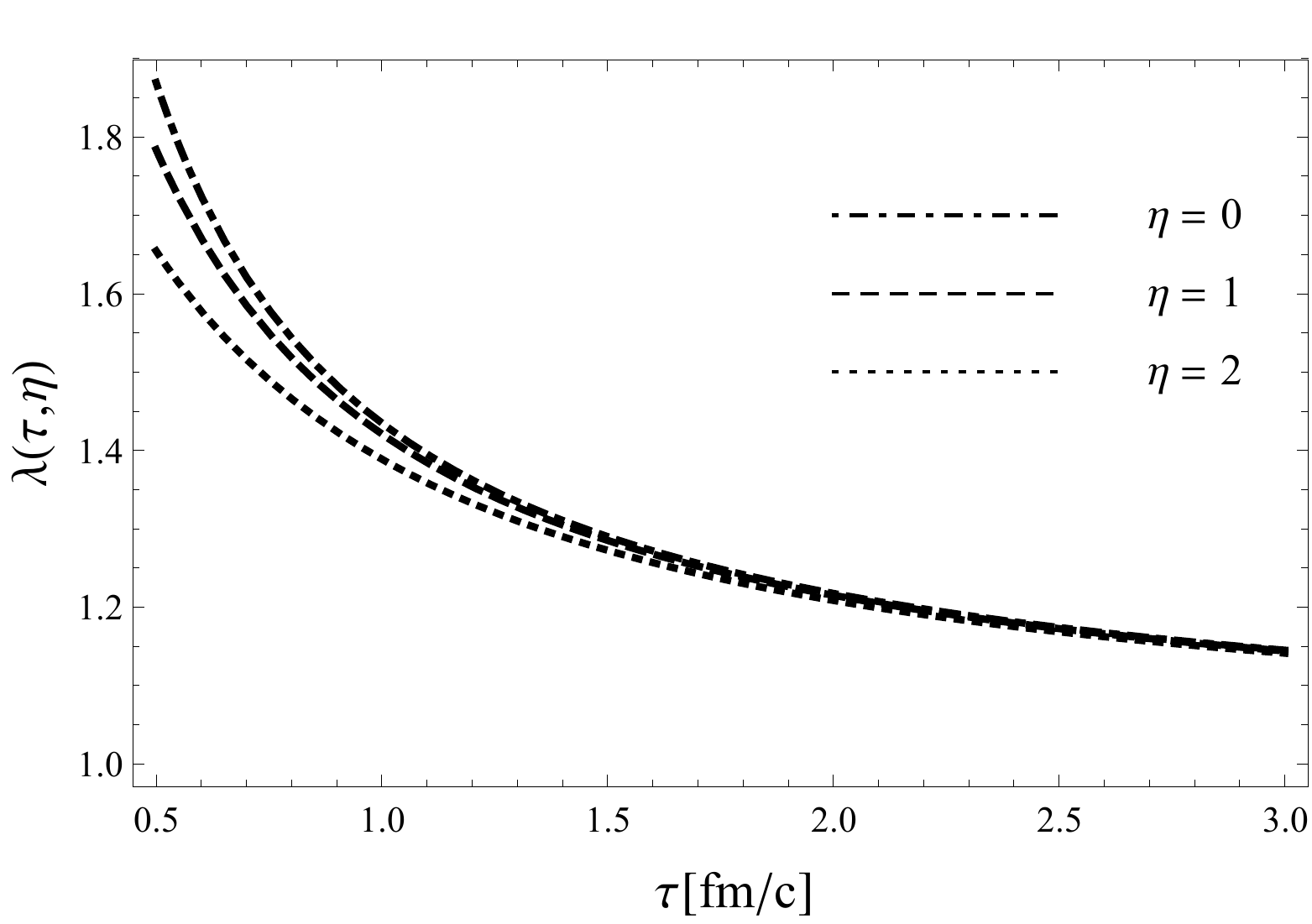}
\end{center}
\caption{Acceleration parameter $\lambda(\tau,\eta)$ in term of proper time $\tau$ for different rapidity. The values $\alpha=0.1$ and $\sigma=0.023$ fm$^{-1}$ are chosen.}
\label{fig:fig1}
\end{figure}

\begin{figure}
\begin{center}
\includegraphics[width=3in]{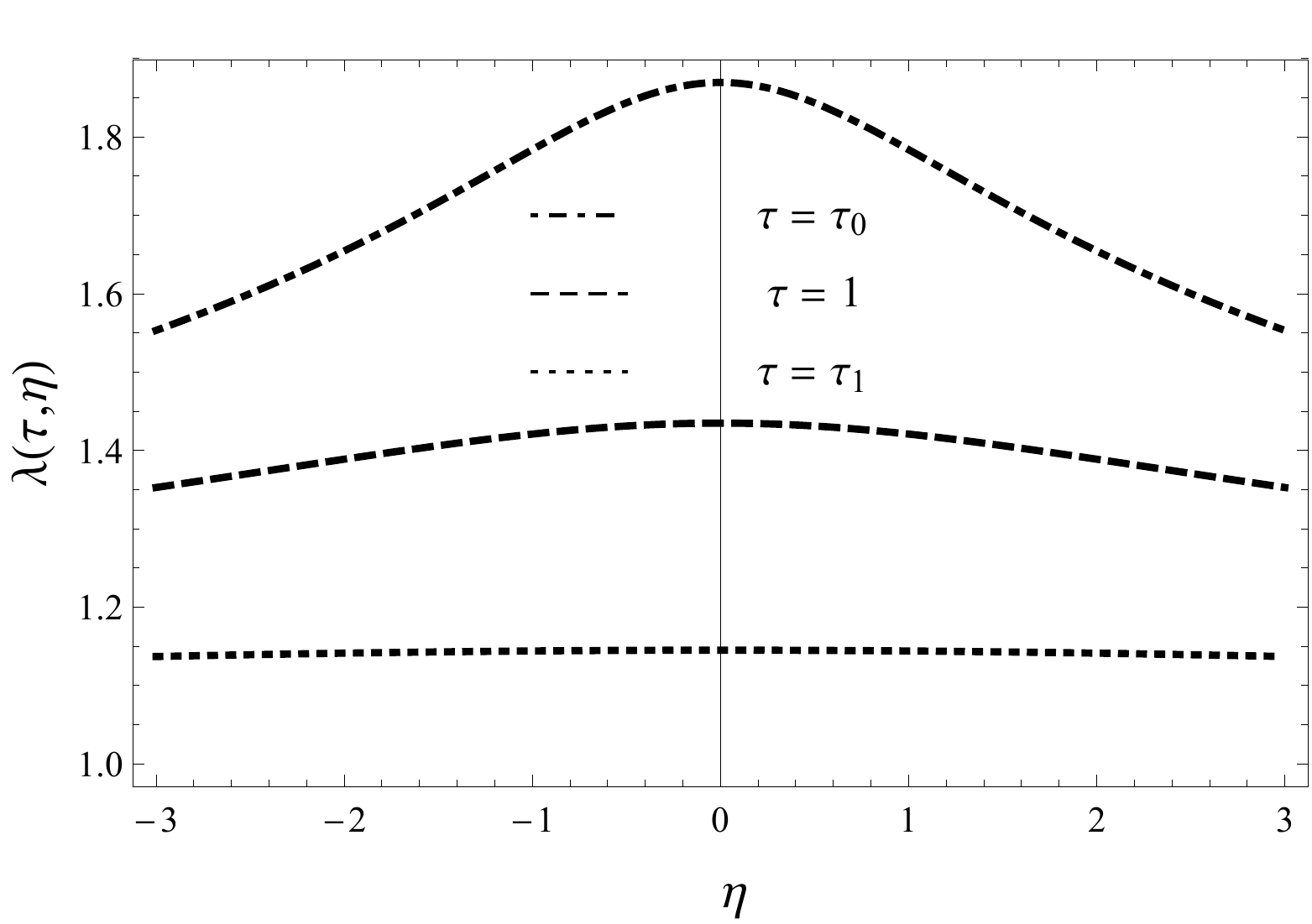}
\end{center}
\caption{Acceleration parameter $\lambda(\tau,\eta)$ in term of rapidity   $\eta$ for different proper time $\tau_0=0.5,\ \tau=1,\ \tau_1=3 fm$.
 The values $\alpha=0.1$ and $\sigma=0.023$ fm$^{-1}$ are chosen.}
\label{fig:fig2}
\end{figure}

\begin{figure}
\begin{center}
\includegraphics[width=3.5in]{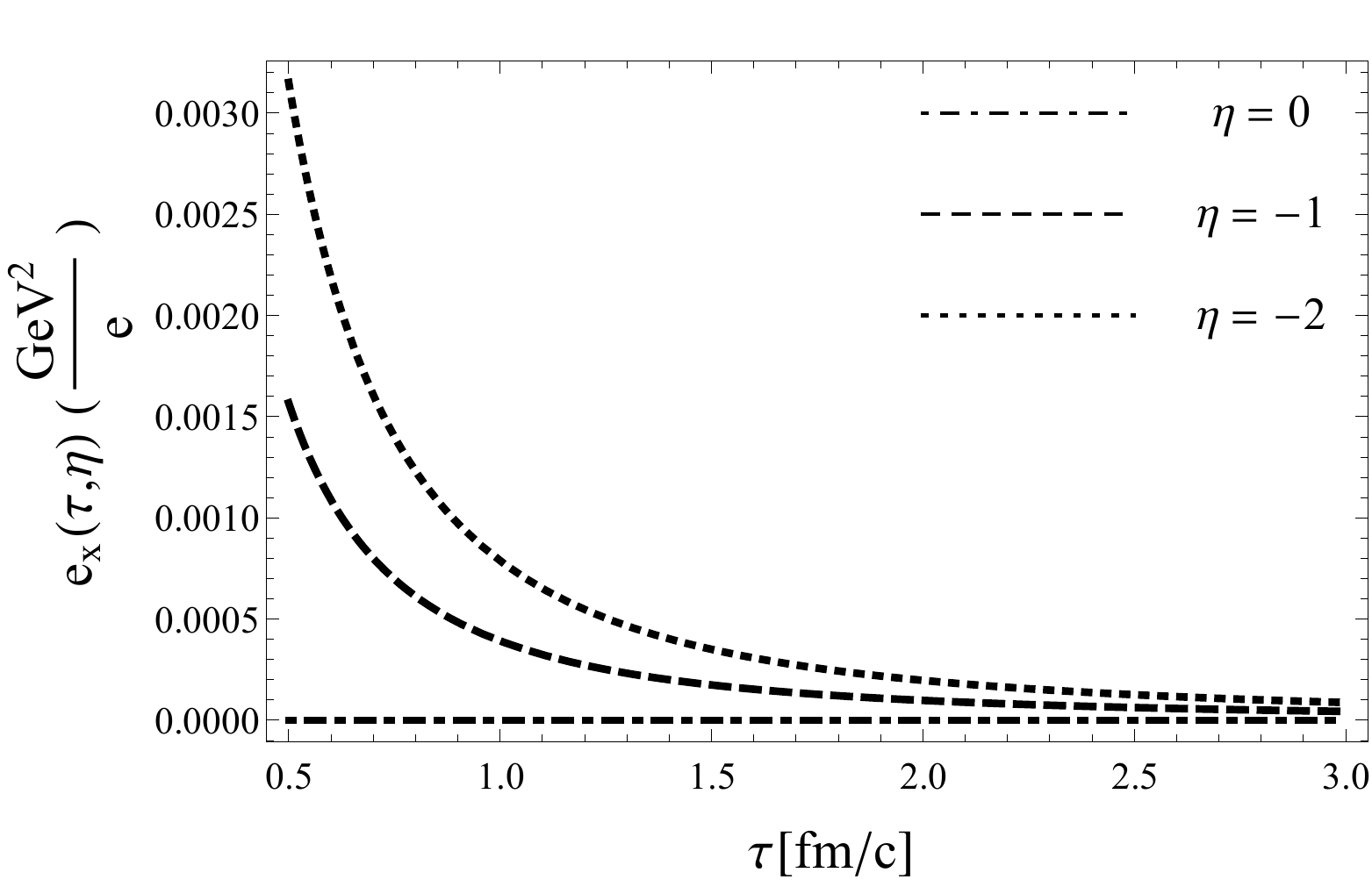}
\end{center}
\caption{Electric field $e_x(\tau,\eta)$ in term of proper time $\tau$ for different rapidities.  The values $\alpha=0.1$ and $\sigma=0.023$ fm$^{-1}$ are chosen.}
\label{fig:fig3}
\end{figure}

\begin{figure}
\begin{center}
\includegraphics[width=3.5in]{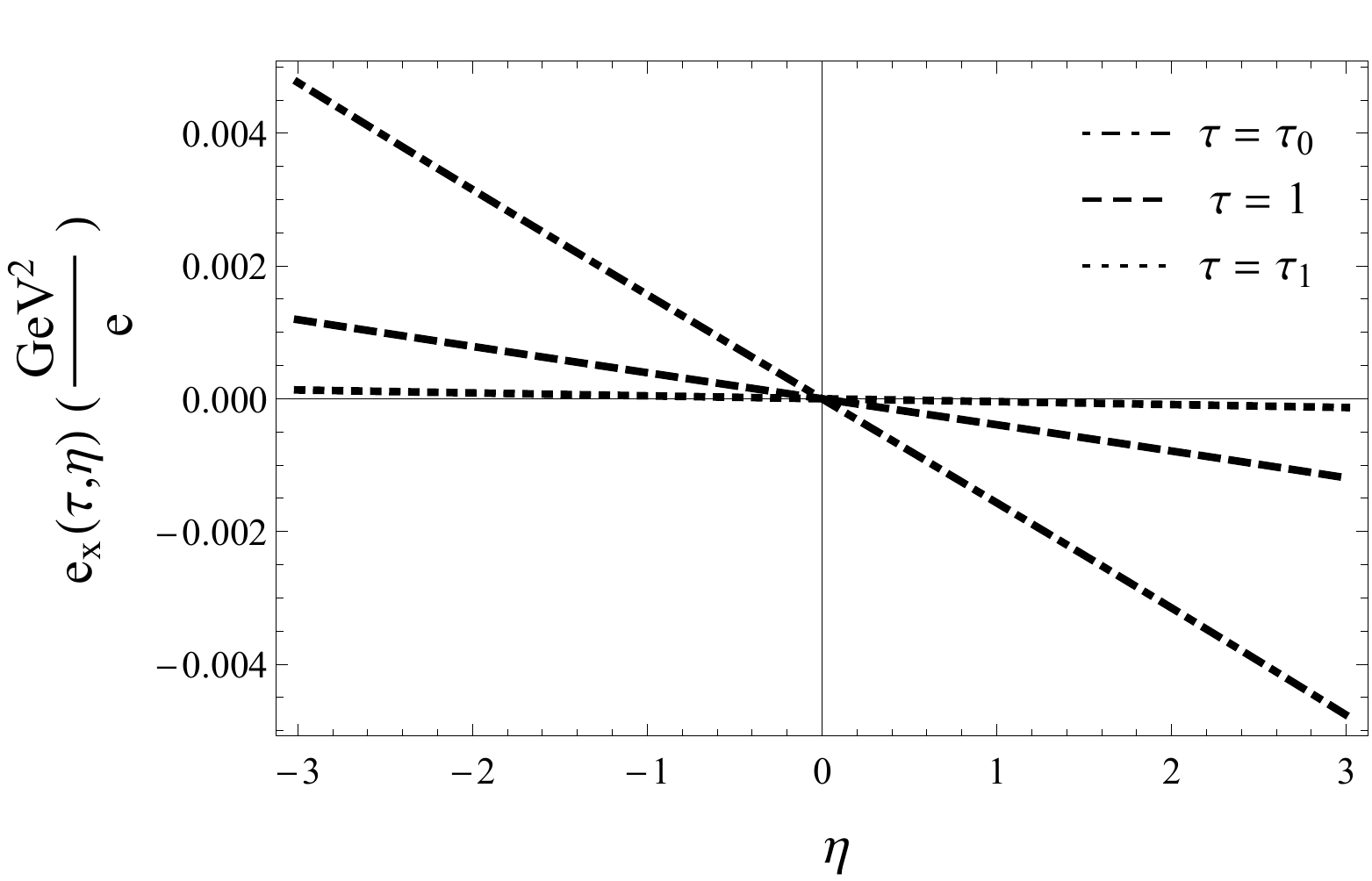}
\end{center}
\caption{Electric field $e_x(\tau,\eta)$ in term of rapidity   $\eta$ for different proper time $\tau_0=0.5,\ \tau=1,\ \tau_1=3 fm$. 
 The values $\alpha=0.1$ and $\sigma=0.023$ fm$^{-1}$ are chosen.}
\label{fig:fig4}
\end{figure}

\begin{figure}
\begin{center}
\includegraphics[width=3.5in]{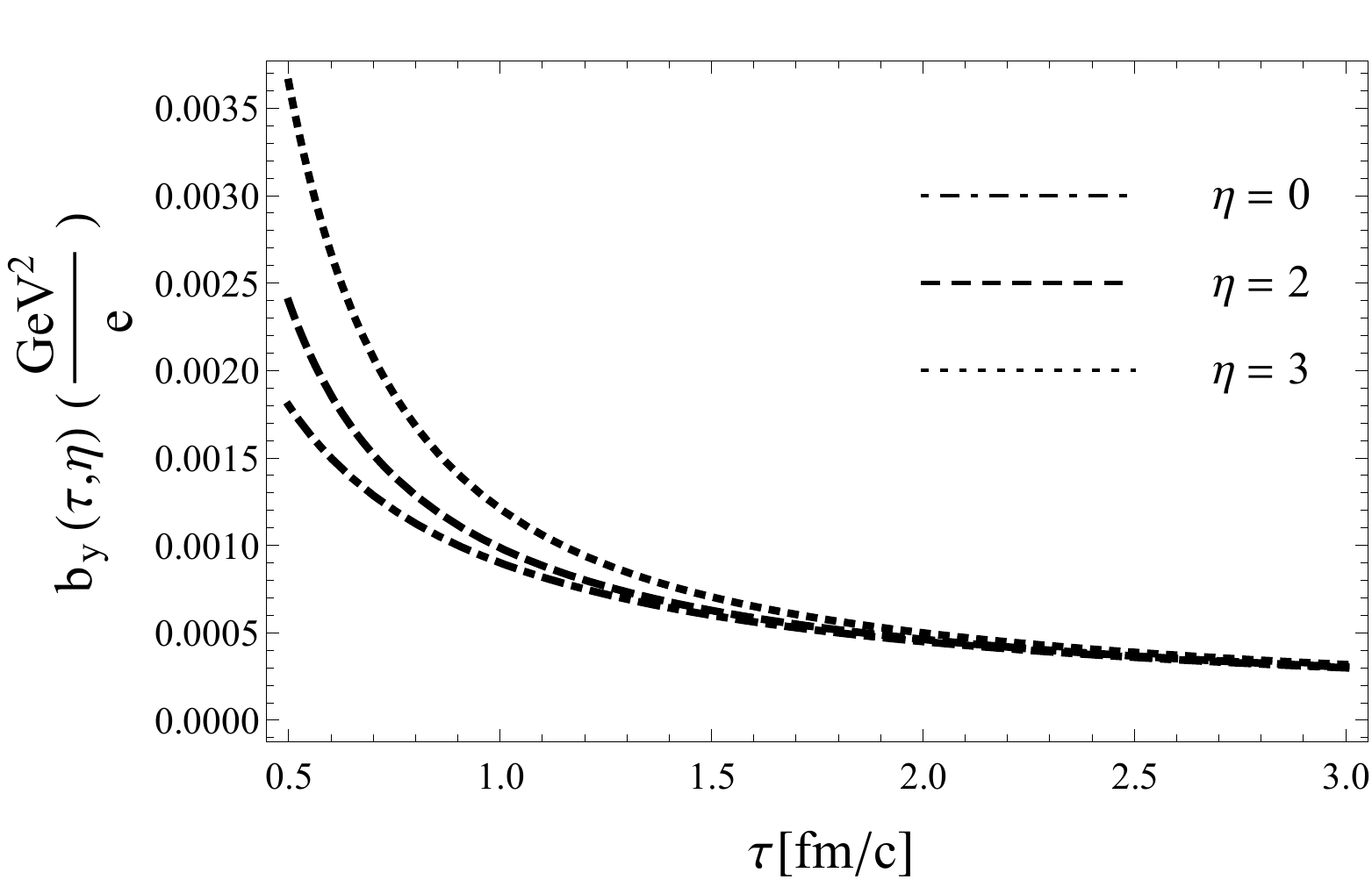}
\end{center}
\caption{Magnetic field $b_y(\tau,\eta)$ in term of proper time $\tau$ for different rapidities.  The values $\alpha=0.1$ and $\sigma=0.023$ fm$^{-1}$ are chosen.}
\label{fig:fig5}
\end{figure}

\begin{figure}
\begin{center}
\includegraphics[width=3.5in]{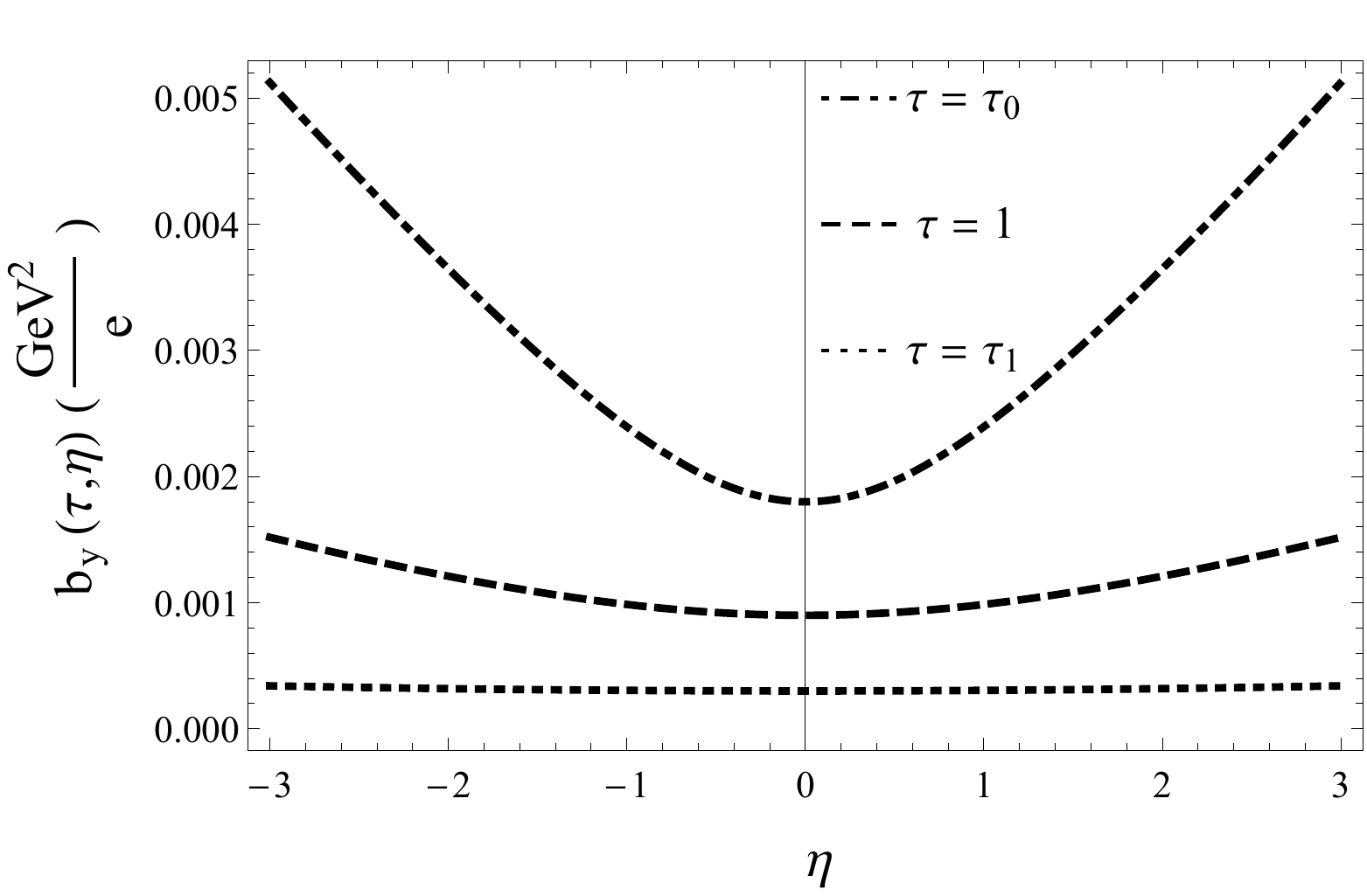}
\end{center}
\caption{Magnetic field $b_y(\tau,\eta)$ in term of rapidity   $\eta$ for different proper time $\tau_0=0.5,\ \tau=1,\ \tau_1=3 fm$. The values $\alpha=0.1$ and $\sigma=0.023$ fm$^{-1}$ are chosen.}
\label{fig:fig6}
\end{figure}

\begin{figure}
\begin{center}
\includegraphics[width=3.5in]{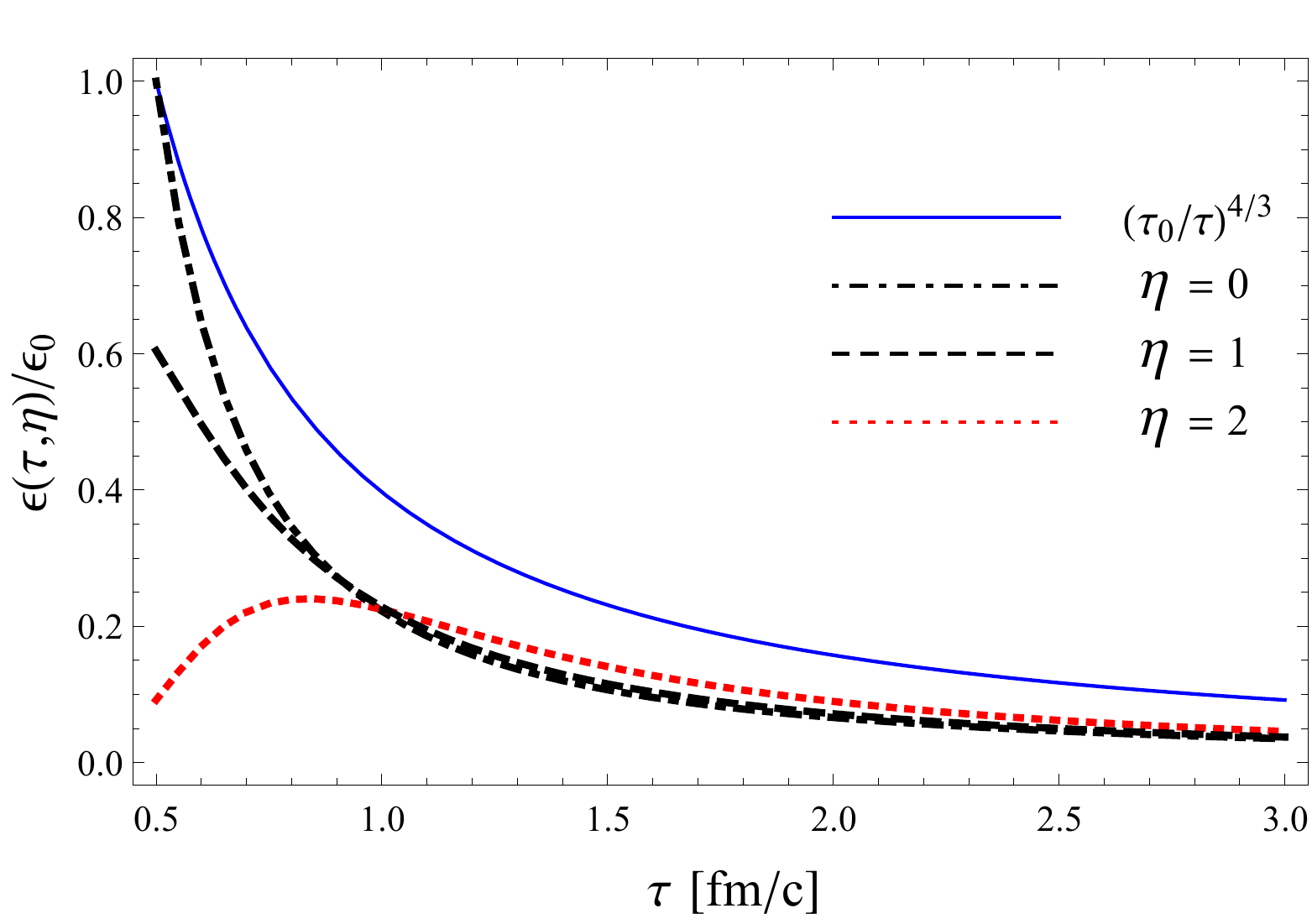}
\end{center}
\caption{The ratio of energy density $\epsilon(\tau,\eta)/\epsilon_0$  in term of proper time $\tau$ for different rapidities and comparison with Bjorken Model (thin continuous line).  The values $\alpha=0.1$ and $\sigma=0.023$ fm$^{-1}$ are chosen.}
\label{fig:fig7}
\end{figure}

\begin{figure}
\begin{center}
\includegraphics[width=3.5in]{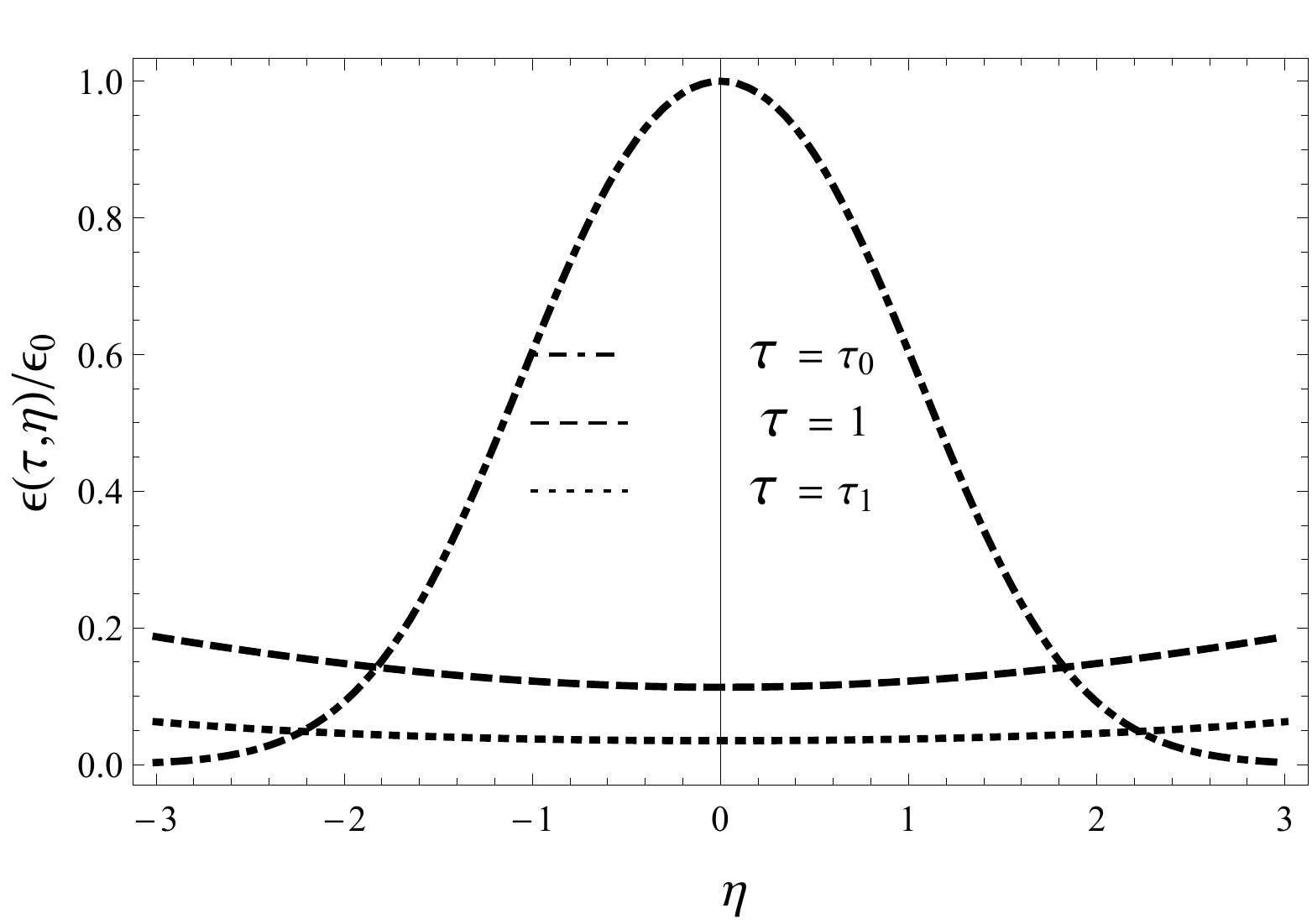}
\end{center}
\caption{The ratio of energy density $\epsilon(\tau,\eta)/\epsilon_0$ in term of rapidity $\eta$ for different proper time $\tau_0=0.5,\ \tau=1,\ \tau_1=3 fm$.  The values $\alpha=0.1$ and $\sigma=0.023$ fm$^{-1}$ are chosen.}
\label{fig:fig8}
\end{figure}

\begin{figure}
\begin{center}
\includegraphics[width=3.5in]{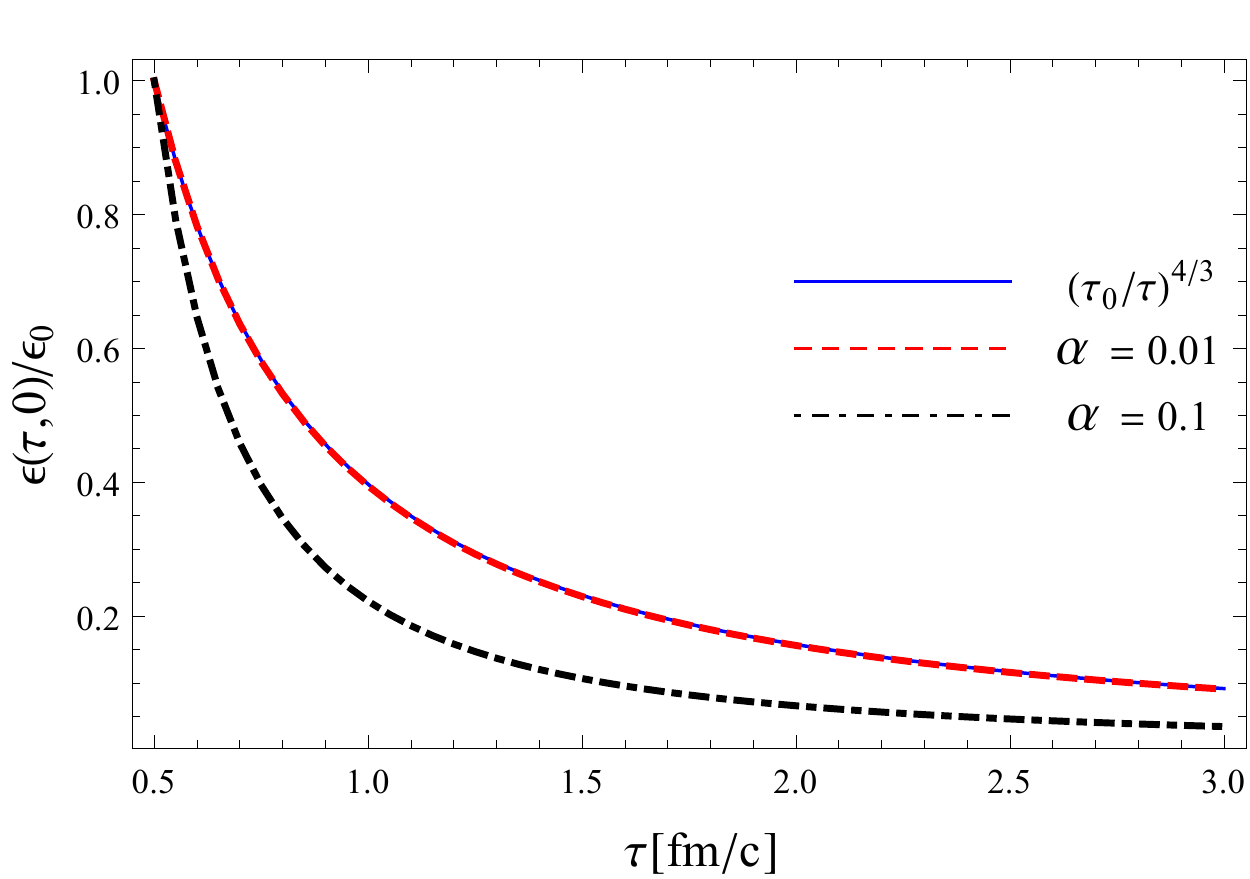}
\end{center}
\caption{The ratio of energy density $\epsilon(\tau,\eta)/\epsilon_0$  in term of proper time $\tau$ at mid rapidity $\eta=0$ for different valuse of $\alpha$ and comparison with Bjorken Model (thin continuous line). The values $\sigma=0.023$ fm$^{-1}$ is chosen.}
\label{fig:fig9}
\end{figure}

\begin{figure}
\begin{center}
\includegraphics[width=3.5in]{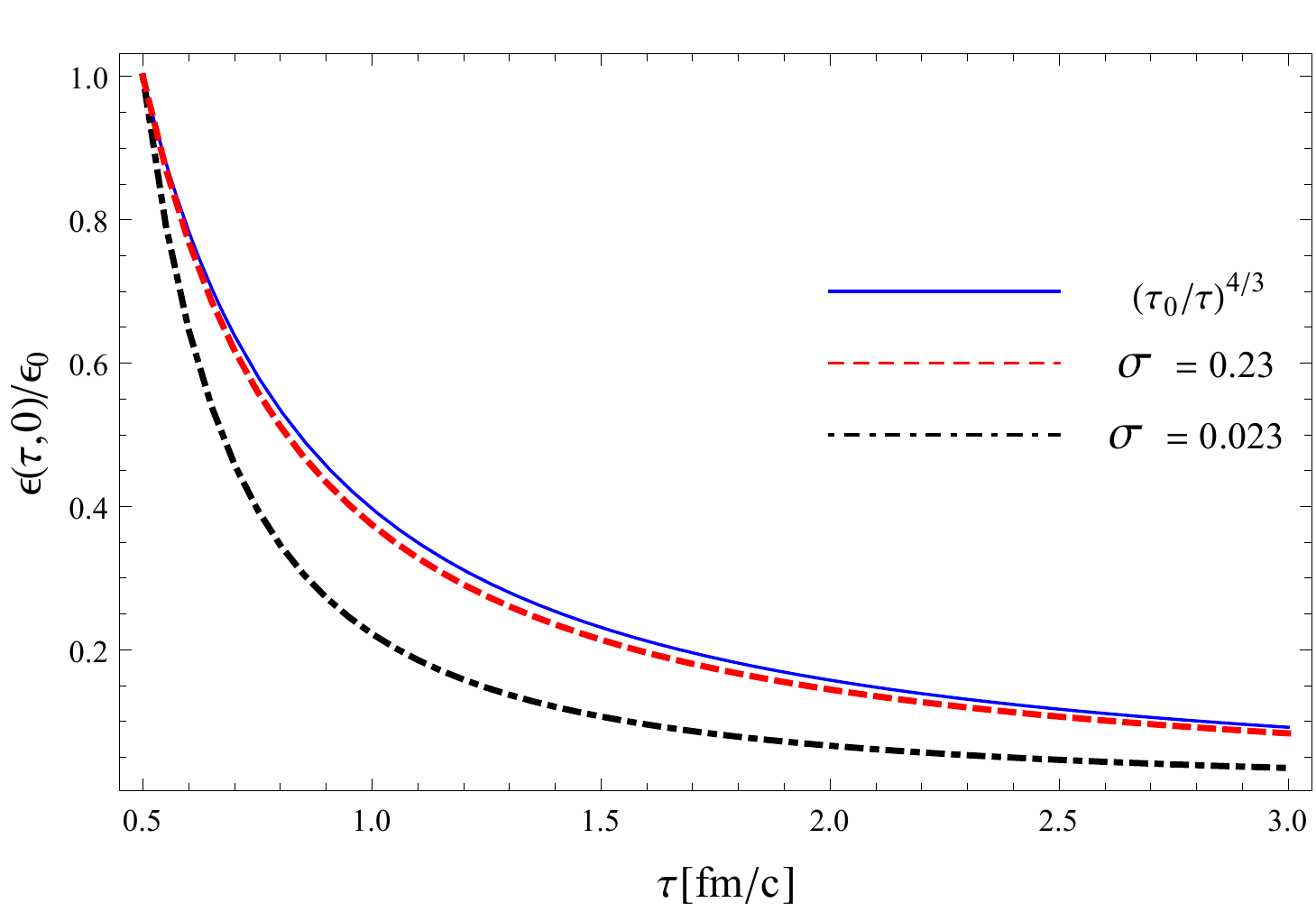}
\end{center}
\caption{The ratio of energy density $\epsilon(\tau,\eta)/\epsilon_0$  in term of proper time $\tau$ at mid rapidity $\eta=0$ for different valuse of $\sigma$ and comparison with Bjorken Model (thin continuous line). 
 The values $\alpha=0.1$ is chosen.}
\label{fig:fig10}
\end{figure}

\begin{figure}
\begin{center}
\includegraphics[width=3.5in]{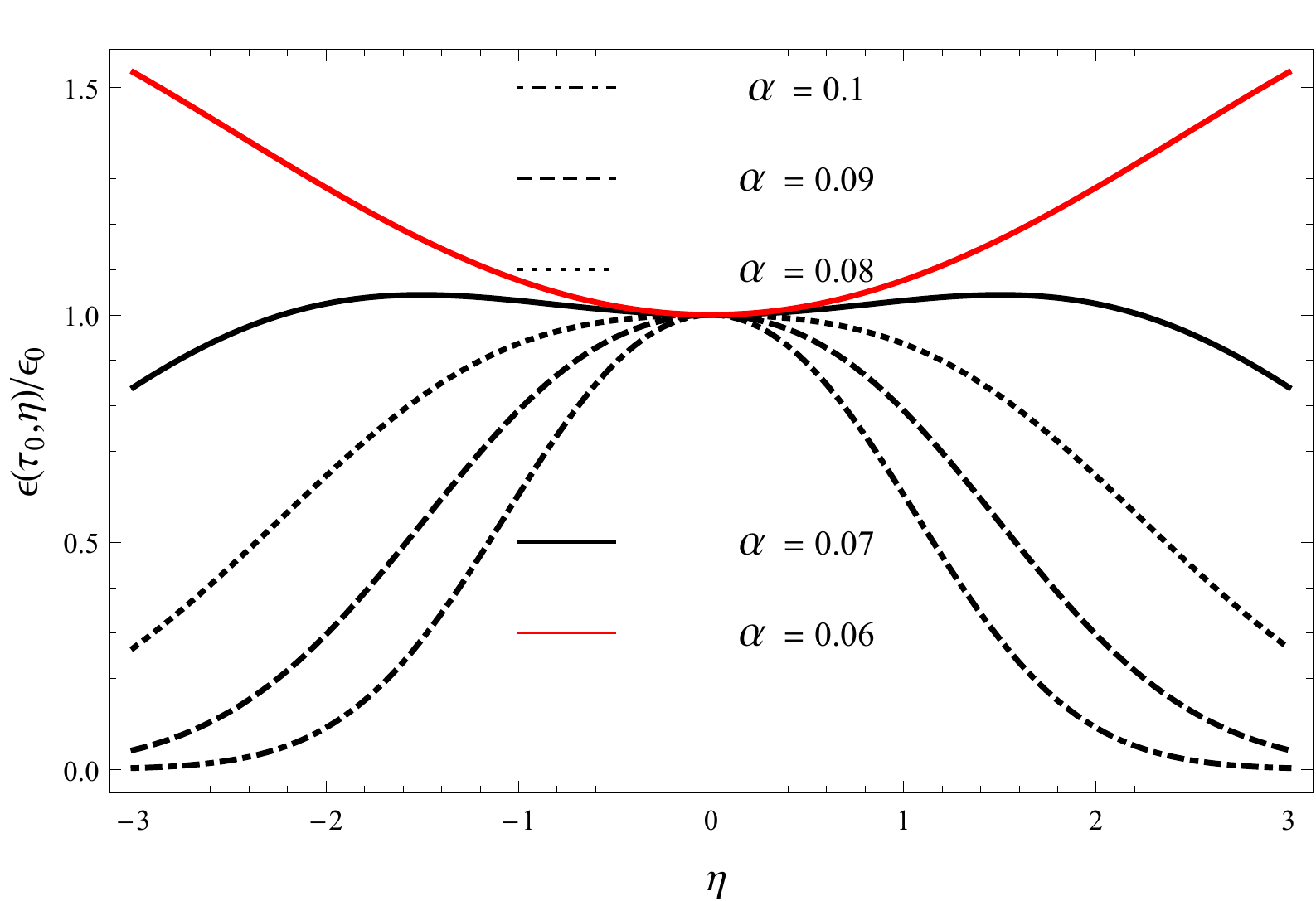}
\end{center}
\caption{The ratio of energy density $\epsilon(\tau,\eta)/\epsilon_0$  in term of rapidity $\eta$ at early proper time $\tau_0=0.5 fm$ for different values of $\alpha$. The value $\sigma=0.023$ fm$^{-1}$ is chosen.}
\label{fig:fig11}
\end{figure}

\begin{figure}
\begin{center}
\includegraphics[width=3.5in]{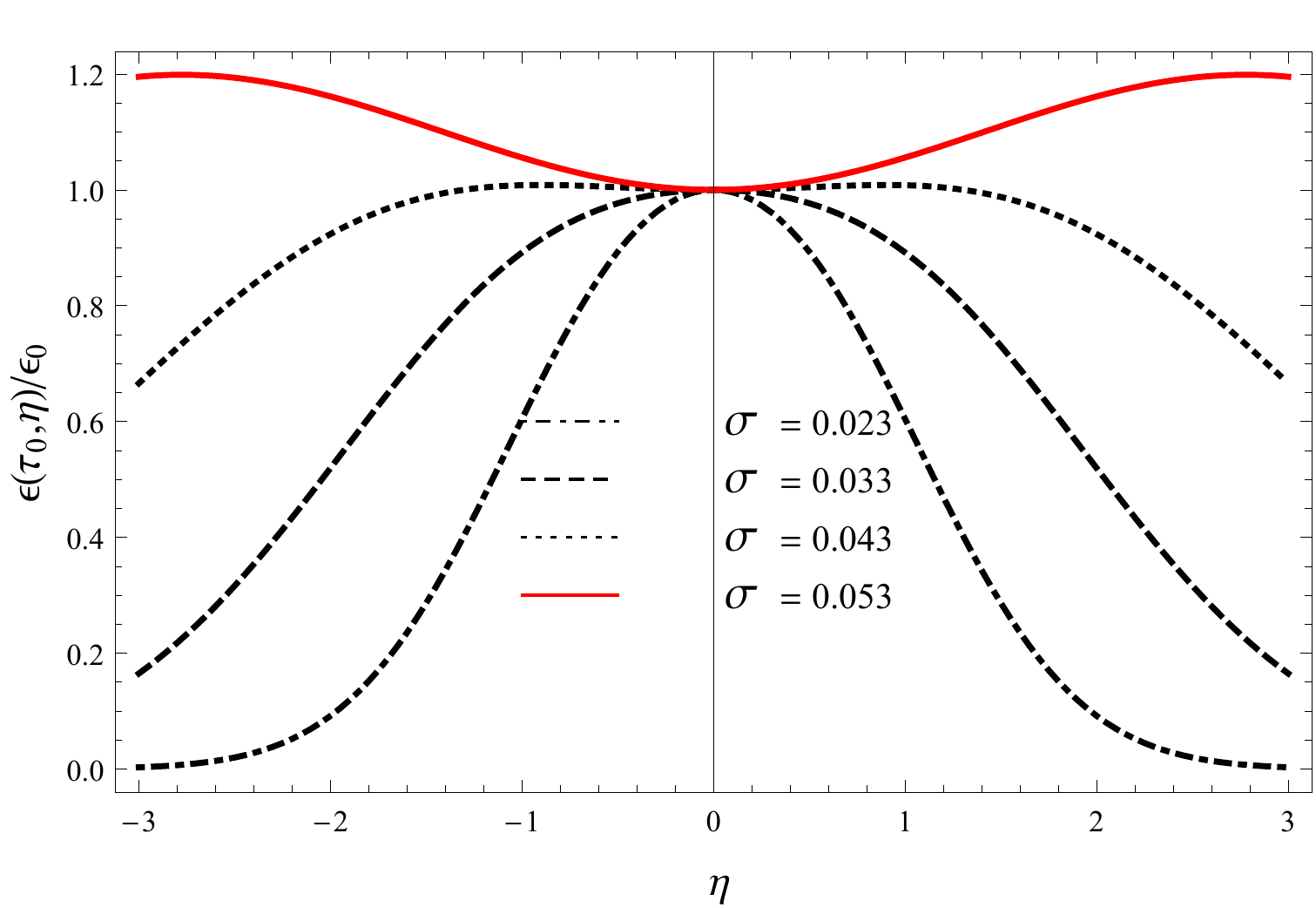}
\end{center}
\caption{The ratio of energy density $\epsilon(\tau,\eta/\epsilon_0)$ in term of rapidity $\eta$ at early proper time $\tau_0=0.5 fm$ for different values of $\sigma$. The value $\alpha=0.1$ is chosen.}
\label{fig:fig12}
\end{figure}

Here we highlight some comments in order to impose further constraints on the solutions:
\begin{itemize}
  \item We  remind that  the relation between fluid rapidity and space time rapidity is $Y=\lambda \eta$, where $\lambda$ is called the acceleration parameter \cite{duanshe2019}. So, we can write
      \begin{eqnarray}
        \lambda &=& 1+\frac{1}{\eta}\sinh^{-1}\left(\frac{1}{\sigma\tau}\frac{\partial_\eta c(\eta)}{c(\eta)}\right).\label{lambda}
      \end{eqnarray}
If we choose a constant value for $c(\eta)=$const, then the flow has no acceleration $\lambda=1$, $(Y\equiv\eta\rightarrow \bar v=0)$, the electric field $e_x$ vanishes and the magnetic field is obtained as $b_y\propto \frac{1}{\tau}$, which is equivalent to the frozen flux theorem; indeed this happens in ideal RMHD, where the electrical conductivity is infinite ($\sigma\rightarrow\infty$).
Then the flow under study will not change and the Bjorken model will be recovered; we will get $\epsilon\propto \tau^{-\frac{4}{3}}$ if $\kappa=\frac{1}{3}$.

From the experimental data, it is found that  $\lambda > 1$, which means that the fireball expansion is fast and a large energy density deposits at mid-rapidity $\eta$.

\item Beside, according to Refs.~(\cite{catania2017}-\cite{gursoy2018})\footnote{In our study, {\it transverse RMHD}, the fluid velocity is perpendicular to EM fields, the proper charge density $\rho$ is zero, because in the aforementioned setup, there is no vorticity: please refer to Appendices in Refs~(\cite{shokri2017}-\cite{inghirami2019}). In this case, the electric field is contributed from the Faraday law and Lorentz force. Indeed, based on the Faraday law, the decreasing magnetic field $b_y$ with time produces an electric field in the direction $x$ and since the fluid has a huge longitudinal velocity, Lorentz force is oriented in $x$ axes. }, the electric field $e_x$ has opposite directions at positive and negative rapidity. Also, at mid-rapidity $\eta=0$, the electric field is zero, but the magnetic field $b_y$ at mid-rapidity has a non-zero value . So, we can consider the function $c(\eta)$, as an even function in term of rapidity but not vanishing at mid-rapidity.  Moreover, since the function $c(\eta)$ has to be considered  as a small deviation from the case of highly conductive plasma, one can assume that the dynamical electromagnetic fields in QGP roughly follows the similar patterns of the external fields created by charged spectators.

\item According to the ansatz in eqs. (\ref{ansatz1}-\ref{ansatz2}), the EM fields in the Lab frame; in Milne coordinate, could be obtained as follows:
    \begin{eqnarray}
      {\bf{ \tilde E}}^{i}_{L}=F^{0 i}=0, \\
      {\bf{\tilde B}}^{i}_{L}=F^{\star 0 i},
    \end{eqnarray}
    We use the lower index L for the EM fields in the laboratory frame and tilde symbol for Milne coordinate. We  find that in the lab frame $\tilde E^x_L=0,$ and $\tilde B^y_L=\frac{c(\eta)}{\tau}$.
    Meanwhile, in Minkowski coordinate\footnote{The transformation of the EM fields from Minkowski coordinates to Milne coordinates is as follows:
    $\tilde E^x=\cosh\eta E^x-\sinh\eta B^y,\
    \tilde E ^y=\cosh\eta E^y+\sinh\eta B^x,\ \tilde E^\eta=\frac{E^z}{\tau},\ \tilde B^x=\cosh\eta B^x+\sinh\eta E^y,\ \tilde B^y=\cosh\eta B^y-\sinh\eta E^x,\ \tilde B^\eta=\frac{B^z}{\tau}$.}, the EM field could be obtained as follow:
    \begin{eqnarray}
      {\bf E}_L &=& (\sinh(\eta)\frac{c(\eta)}{\tau}, 0, 0),\label{exlab}\\
       {\bf B}_L &=& (0, \cosh(\eta)\frac{c(\eta)}{\tau}, 0)\label{bylab}
    \end{eqnarray}
    As we can observe the electric field $\tilde {\bf E}=0$ in the lab frame of Milne coordinates is zero but
     in the lab frame of Minkowski coordinates ${\bf E}\neq 0$, is not zero.

     Now, we try to show the self consistence of our solutions in the lab frame.
    From Faraday Law
    \begin{eqnarray}
      \nabla\times{\bf E}_L &=& -\partial_t {\bf B}_L,
    \end{eqnarray}
One can check the self-consistence of Maxwell's equations. It is found that with eqs. (\ref{exlab}-\ref{bylab})
the $\partial_y E^x_L=\partial_t B^z_L=0$ and $\partial_z E^x_L=-\partial_t B^y_L$ are automatically satisfied. Similarly, $\nabla\cdot {\bf E}_L=\rho$ and $\nabla\cdot {\bf B}_L=0$ are satisfied with $\rho=0$.

The last Maxwell's equation is
\begin{eqnarray}
  \nabla\times {\bf B}_L &=& {\bf j}+\partial_t {\bf E}_L,\\
   {\bf j}&=&\sigma\gamma({\bf E}_L+{\bf v}\times{\bf B}_L)
\end{eqnarray}
Where ${\bf v}$ is the three-vector fluid velocity; $u^\mu=\gamma(1,{\bf v})$. It is clear that in our setup, $\partial_x B^y_L=0$ and $-\partial_z B^y_L=\sigma\gamma(E^x_L-v_z B^y_L)+\partial_t E^x_L$ are satisfied. As we observe the relativistic induced current ${\bf j}$ is along the $x$ axis.
\end{itemize}

Finally, taking into account the above considerations, we can model $c(\eta)$ by taking a simple function as follows:
\begin{eqnarray}
 c(\eta)&=&c_0\cosh(\alpha\eta),
\end{eqnarray}
with small values of $\alpha$.
In order to fix the constant, $c_0$, we consider the initial condition for magnetic field at mid rapidity in the lab frame, which is $B_L^y(\tau_0,0)=0.0018 \frac{GeV^2}{e}$ \cite{gursoy2018} while the coefficient $\alpha$ is selected as an arbitrary small value  in order to parameterize the acceleration parameter  $\lambda$. Later we discuss the effect of variations in $\alpha$ and $\sigma$.

The next step is to solve the conservation equations; the main idea for solving eqs. (\ref{eq1}, \ref{eq2})  is to change these
two couple partial differential equations (PDEs) into two ordinary differential equations (ODEs) with a given initial condition $\epsilon(\tau_0, 0)=\epsilon_0$ \cite{duanshe2019}.
The combination of the energy and Euler equations (\ref{eq1}, \ref{eq2}) can be rewritten as follows:
\begin{eqnarray}
  \partial_\tau \epsilon(\tau,\eta)+\frac{1+\kappa}{\tau}A(\tau,\eta) \epsilon(\tau,\eta) &=& B(\tau,\eta)\label{eqe1} \\
  \partial_\eta \epsilon(\tau,\eta)+H(\tau,\eta) \epsilon(\tau,\eta) &=& G(\tau,\eta), \label{eqe2}
\end{eqnarray}

where $\kappa=\frac{1}{3}$ is considered and
\begin{eqnarray}
A(\tau,\eta)&=& \Big(\frac{\partial_\eta Y (\bar v^2-\kappa )-(\kappa -1) \tau  \bar v \partial_\tau Y}{\kappa  \left(\bar v^2-1\right)})\Big) \\
  B(\tau,\eta) &=& \frac{\sigma   (e_xb_y \bar v-\kappa  e_x^2)}{\kappa  \bar\gamma  \left(\bar v^2-1\right)}  \\
  H(\tau,\eta) &=&\frac{1}{\kappa}\Big((1+\kappa)(\tau\partial_\tau Y+\bar v\partial_\eta Y)\Big)-(1+\kappa)\bar v A(\tau,\eta)  \\
  G(\tau, \eta) &=& \frac{(\sigma\tau)e_x b_y}{\bar\gamma\kappa}-\tau \bar v B(\tau,\eta)
\end{eqnarray}

We can think of solving this coupled system of PDEs on a grid of points in the $(\tau,\eta)$ plane. First we move along the $\tau$ direction and solve eq. (\ref{eqe1}) to find out the $\tau$-dependence of the function $\epsilon$, keeping constant the variable $\eta$. In this step we treat eq. (\ref{eqe1}) as an ODE with respect to $\tau$. Then we move along the $\eta$ direction, keeping the solution $\epsilon(\eta)$ previously found as the initial condition for solving (as an ODE) eq. (\ref{eqe2}), 
Finally we obtain numerically the full energy density profile.


\section{Discussion and conclusion}
From pure analytical solutions we found that, by considering a finite electrical conductivity,  the longitudinal evolution of the fluid subject to the presence of EM fields will accelerate. 
 Fig. (\ref{fig:fig1}) shows the acceleration parameter $\lambda(\tau, \eta)$, in terms of $\tau$ for different values of rapidity. As one expects the acceleration parameter decreases with time but remains larger than 1 up to very late times. In Fig. (\ref{fig:fig2}) the acceleration parameter is depicted in term of rapidity for fixed $\tau$. In both forward and backward rapidity, by increasing the $|\eta|$, the acceleration parameter decreases.
 
 Next we investigated on the dynamical evolution of EM fields.
 In Figs. (\ref{fig:fig3}) and (\ref{fig:fig4}) $e_x(\tau, \eta)$, is displayed , at either fixed $\eta$ or fixed $\tau$, respectively. From Fig. (\ref{fig:fig3}) one finds that following the time evolution, the electric field decays and at late times is very small. In Fig. (\ref{fig:fig4}) one can see that, in central rapidity, after the QGP is formed, the electric field is zero and  no external electric fields is left in average. However the electric field is stronger at large rapidities since it is generated by the close nucleus. As one observes, the electric field has opposite signs for forward and backward rapidity, being an odd function of $\eta$.
 
 Figs. (\ref{fig:fig5}) and (\ref{fig:fig6}) show $b_y(\tau, \eta)$ in terms of $\tau$ for several values of rapidity and vice versa, respectively. Similarly to the electric field, the magnetic field decreases with increasing time at all rapidities and increases, at fixed time, with rapidity. However, at variance with the electric field, the magnetic one has a non-zero value at central rapidity. \cite{gursoy2018}.
 
 The ratio of energy density $\epsilon(\tau, \eta)/\epsilon_0$, in terms of $\tau$ for several values of rapidity is illustrated in Fig. (\ref{fig:fig7}). As tha fluid expands, the energy density decays. In this figure, we compare the accelerated fluid with the Bjorken model. One can find that the decay rate for the accelerated plasma is faster than for the Bjorken fluid. In Fig. (\ref{fig:fig8}), the $\epsilon(\tau, \eta)/\epsilon_0$ as a function of rapidity for fixed $\tau$ is plotted. Naively it seems that at the early time, when QGP is formed, the $\eta$ profile for  the $\epsilon(\tau, \eta)/\epsilon_0$, has a Gaussian form,  while at the late time it becomes rather a plateau, in agreement with \cite{bozek2008}. It is found that energy density slowly flows toward high rapidity at the later time.
 
 In order to discuss the dependence of the model on the parameters $\alpha$ and $\sigma$, we concentrate on the quantity $\epsilon(\tau, \eta)/\epsilon_0$ and discuss both its time and rapidity evolutions with different choiches of the above mentioned parameters.
 Figs. (\ref{fig:fig9}) and (\ref{fig:fig10}) show the time evolution of $\epsilon(\tau, \eta)/\epsilon_0$ in the mid rapidity, $\eta=0$, at either fixed $\sigma$ or fixed $\alpha$, respectively. From Fig. (\ref{fig:fig9}), the smallest value of $\alpha$ brings the present model close to the Bjorken Model but for larger values of $\alpha$, which imply a highly accelerated fluid, the rate of decay for $\epsilon(\tau, \eta)/\epsilon_0$ would be faster. Beside, in Fig. (\ref{fig:fig10}), for higher values of $\sigma$, the $\epsilon(\tau, \eta)/\epsilon_0$ is compatible with the Bjorken flow, while for small values of the electrical conductivity, the decay rate for the expanding fluid is fast.
 
 Finally, in  Figs. (\ref{fig:fig11}) and (\ref{fig:fig12}), a similar discussion is presented for the rapidity profile of $\epsilon(\tau, \eta)/\epsilon_0$ at the fixed proper time $\tau_0=0.5$ fm. Fig. (\ref{fig:fig11}), where $\sigma$ is fixed, shows that by decreasing $\alpha$, the rapidity profile (at the QGP formation proper time) evolves toward a plateau. But if $\alpha\le 0.06$ the $\epsilon(\tau_0, \eta)/\epsilon_0$ becomes divergent at high rapidity. Analogous behavior can be found in Fig. (\ref{fig:fig12}), where $\alpha$ is fixed: indeed if we increase the value of $\sigma$, the  rapidity profile $\epsilon(\tau_0, \eta)/\epsilon_0$ tends to a plateau, but for $\sigma > 0.053$ fm$^{-1}$ it will not be flat. As we observe,  $\alpha$ has a diverse role with respect to $\sigma$. Also, it appears that in a realistic situation there is a lower limit for $\alpha$ and an upper limit for  $ \sigma$.
 
 In the RRMHD framework we found that the Bjorken flow is generalized and the longitudinal expansion of the magnetized QGP is accelerated. The acceleration parameter, $\lambda$, directly depends upon the inverse of the proper time, $\tau$, and of the electrical conductivity of the matter. Clearly, when the system expands, the acceleration will decrease.
 As expected, we found that when the electrical conductivity is high, then the acceleration of the fluid is negligible, but in a finite range for $\sigma$, acceleration will be present and sizeable. It is interesting to keep in mind that $\sigma$ is proportional to the temperature of the QGP matter \cite{ding2011}.
 Since the initial temperature increases with larger center-of-mass energy, $\sqrt{S_{NN}}$, one deduces that the acceleration decreases with increasing center-of-mass energy. This result is in agreement with Ref. \cite{Csanad2018}. These authors conclude that  the acceleration (Longitudinal) is the largest in
central collisions, and it decreases with increasing center of mass energy. In the present work we can not find a direct connection between the acceleration and the centrality of the collisions, since the latter implies a transverse width of the material, while we are only considering the longitudinal flow.

In our solutions, the dynamical evolution of EM fields decays with time and is proportional to the inverse of $\sigma$. Indeed, in the limit of infinite $\sigma$, the electric field $e_x$ is zero and the magnetic field $b_y\propto \frac{1}{\tau}$, in agreement with  previous results \cite{roy15}. The EM fields will be stronger at high rapidity, since it is generated by the close nucleus. Indeed, in our setup, we don't have any net charge for the ideal fluid, so, the origin of the EM fields are the charged spectators flying away along the beam directions. 

The energy density decays with time and  is not flat with $\eta$  at early times, when QGP is formed, although a kind of Gaussian form for the initial energy density was proposed by \cite{eskola1998, bozek2009, bozek2008}. Regarding to relation with electrical conductivity, with decreasing the electrical conductivity the decay rate for the energy density increases. 

In order to justify the selected model for the even function $c(\eta)$, we notice that if $c(\eta)=c_0$ is constant, then the fluid is not accelerated and Bjorken flow is preserved. Hence we deviated from this simple scenario, by adding some  polynomials terms to it as perturbation:
\begin{eqnarray}
c(\eta)=c_0(1+\frac{\alpha^2}{2!} \eta^2+ ...)
\end{eqnarray}
where $\alpha$ must be small. In a rapidity interval $-3\leq\eta\leq3$, and with an upper bound limit of $\alpha\leq 0.1$, then one can approximate the $c(\eta)$ as follows:
\begin{eqnarray}
c(\eta)=c_0 \cosh(\alpha\eta),
\end{eqnarray}
which is the Ansatz used in the present model.

The present work can be used to validate future numerical work in the context of RRMHD in heavy ion collisions. The parameter $\kappa$ (ratio of pressure to energy density) can also be changed simultaneously with the parameters  $\alpha$ and $\sigma$, and in fact the aforementioned model can be studied with various  equations of state.
\section*{Acknowledgments}
M. Haddadi Moghaddam thanks F. Yagi, A. Beraudo and A. De Pace, G. Inghirami, M\'{a}t\'{e} Csan\'{a}d and Tam\'{a}s Cs\"{o}rg\'{o} for fruitful discussions and useful suggestions. M. Haddadi Moghaddam also gratefully acknowledges U. G\"{u}rsoy, E. Marcus for sharing their data. Duan She is supported by the National Natural Science Foundation of China (NSFC) under Grant Nos. 11735007, 11890711, China Scholarship Council (CSC) Contract No.201906770027.

\end{document}